\documentclass[twocolumn]{aastex61}
\pdfoutput=1 
\usepackage{amsmath,amstext}
\usepackage[T1]{fontenc}
\usepackage{apjfonts} 
\usepackage[figure,figure*]{hypcap}

\shorttitle{Survey of Orion Region}
\shortauthors{Tremblay, C.D.}

\begin{document}

\title{A Molecular Line Survey around Orion at Low Frequencies with the MWA}

\author{Tremblay, C.D}
\affiliation{International Centre for Radio Astronomy Research, Curtin University, GPO Box U1987, Perth WA 6845, Australia}
\author{Jones, P. A.}
\affiliation{School of Physics, University of New South Wales, Sydney, NSW 2052, Australia}
\author{Cunningham, M.}
\affiliation{School of Physics, University of New South Wales, Sydney, NSW 2052, Australia}
\author{Hurley-Walker, N.}
\affiliation{International Centre for Radio Astronomy Research, Curtin University, GPO Box U1987, Perth WA 6845, Australia}
\author{Jordan, C.H.}
\affiliation{International Centre for Radio Astronomy Research, Curtin University, GPO Box U1987, Perth WA 6845, Australia}
\author{Tingay, S.J.}
\affiliation{International Centre for Radio Astronomy Research, Curtin University, GPO Box U1987, Perth WA 6845, Australia}
\begin{abstract}
The low-frequency sky may reveal some of the secrets yet to be discovered and until recently, molecules had never been detected within interstellar clouds at frequencies below 700\,MHz \citep{Lovas}.  Following the pilot survey towards the Galactic Centre at 103--133\,MHz with the Murchison Widefield Array \citep{Tremblay17}, we surveyed 400\,deg$^{2}$ centered on the Orion KL nebula from 99--170\,MHz.  Orion is a nearby region of active star formation and known to be a chemically rich environment. In this paper, we present tentative detections of nitric oxide and its isotopologues, singularly deuterated formic acid, molecular oxygen, and several unidentified transitions.   The three identified molecules are particularly interesting as laboratory experiments have suggested these molecules are precursors to the formation of amines. 
\end{abstract}


\section{Introduction}

New low-frequency aperture arrays in radio quiet environments, like the Murchison Widefield Array (MWA), allows for simultaneous observations of thousands of stars and star forming regions in a single field-of-view (FOV).  In \cite{Tremblay17}, we completed a pilot survey of the Galactic Centre to determine the practicality of using the MWA to observe spectral lines.  In a blind survey of 400\,deg$^{2}$, we tentatively detected transitions of nitric oxide and the mercapto radical in evolved stars. However, based on the detection limits for the column densities of other potential transitions within the band, it was concluded that observing a region at a distance of 400\,pc (in comparison to the Galactic Centre that is 8600\,pc away) would reduce the beam dilution by an order of magnitude.  The increased sensitivity would potentially allow for detections of even more molecules.

The Orion molecular cloud represents a nearby (414$\pm$7\,pc \citealt{Menten07}) active star formation region whose population of stars, young and old, are used for studying star formation and high-mass stars \citep{ODell15}.   The proximity and complex structure of Orion make it both a critical testing ground for  theoretical models and for the interpretation of results from more distant star forming regions \citep{Muench08}.

In a single survey of the Orion Kleinmann-Low Nebula (Orion KL) from 80 to 280\,GHz using the IRAM 30\,m telescope, approximately 15,400 spectral features were detected with 11,000 identified from 50 different molecular species \citep{Tercero15}. Although these observations were completed at much higher frequencies, they demonstrates the chemically rich nature of the region. As most of the interstellar molecules so far detected can be found in the Orion molecular cloud complex \citep{Baudry-2016}, it is an ideal location for a spectral line survey with a new telescope. 

Currently our understanding of the formation of high-mass stars is limited \citep{Tan14} due to the dust opacity of the H{\sc ii} region surrounding the star and the significant overlap of spectral features observed at high-radio frequencies.  \cite{Codella14} suggested that the observation of these regions at low-frequencies may be key to disentangling the complex molecular structure as the spacing between transitions increases.  Also, the low-energy transitions of long chain molecules reside at low radio frequencies.

High-frequency observations made with telescopes such as IRAM 30\,m mainly detect thermal transitions.  However, at low frequencies the detected transitions are mainly from emission that is directly related to the number of atoms in a given state, making the detection of boosted emission a more likely scenario \citep{SpitzerISM, SalgadoI}.  This has the potential of yielding new understanding of this complex environment. 

This paper details the molecular line survey centered on the Orion KL, the brightest object within the Orion molecular cloud \citep{Neill-2013}, and the results of these candidate detections made with the MWA.  

\section{Observations \& Data Reduction}
\begin{figure*}
	 \includegraphics[width=1\textwidth]{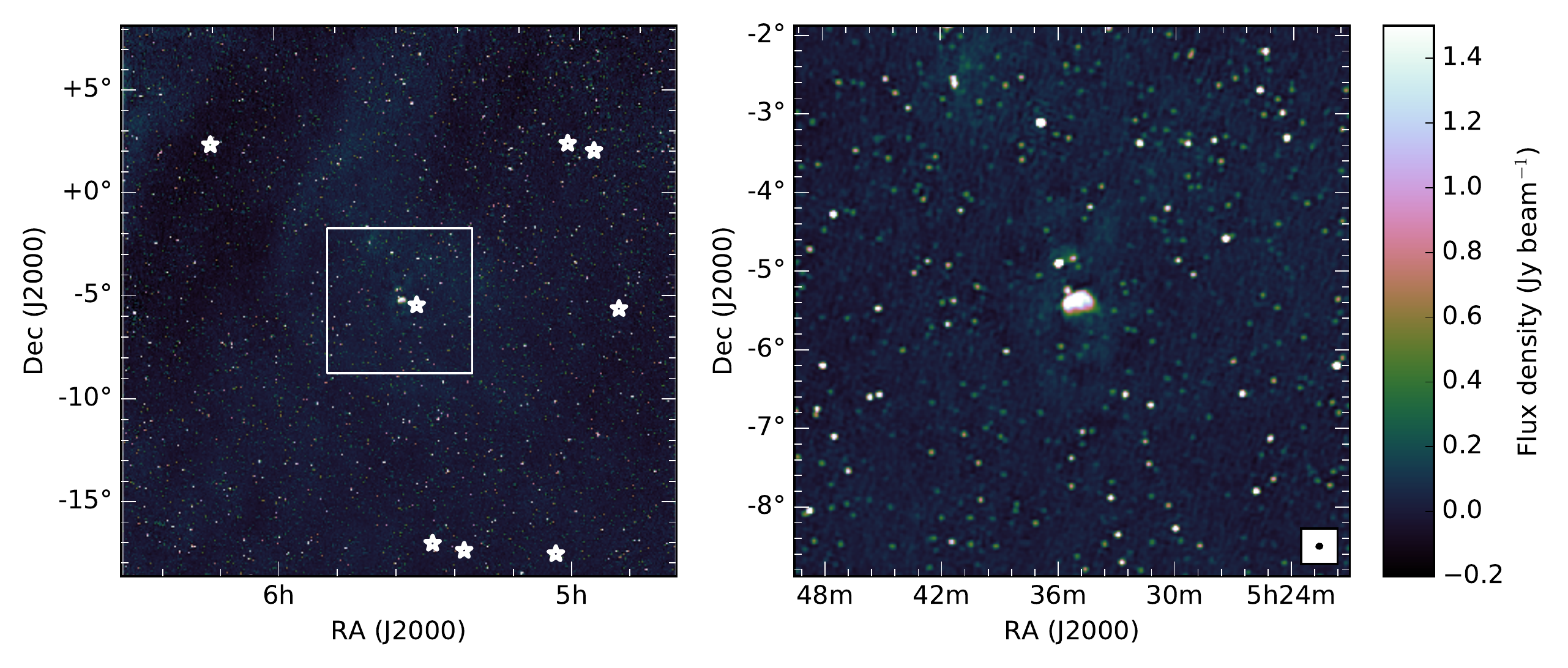}
	\caption{Continuum image of the Orion survey region MWA across the 30.72\,MHz of bandwidth with a central frequency of
114.6 MHz. The left hand image represents the full region blindly searched for molecular signatures (see Section 4).  The right-hand image is a zoom of the region showing the Orion KL nebula in the center of the observed region. The white stars show the positions of the tentative detections reported here.}
\label{cont}
\end{figure*}

The observations were carried out using the Murchison Widefield Array (MWA, \citealt{Tingay13}) located at the Murchison Radio-astronomy Observatory, on 2015 November 21st and 22nd. The observations of the Orion region, centered on  05$^{h}$35$^{m}$17.3$^{s}$ -05$^{\circ}$23$^{\prime}$28$^{\prime\prime}$ (Figure \ref{cont}), were broken up into two continuous frequency bands with central frequencies of 114.6 and 155.5\,MHz.  At the time of these observations the telescope had 128 dipole tiles, spread across 3\,km with a primary beam of 30\,degrees full-width-at-half-maximum (FWHM) and a synthesized beam of 2.1\,arc\,minutes FWHM at 150\,MHz.

The MWA uses a two-stage polyphase filter bank to channelize the data. The first stage separates the 30.72\,MHz bandwidth into 24$\times$1.28\,MHz coarse channels while the second stage breaks up each coarse channel to 128$\times$10\,kHz fine channels.  For this survey, the velocity resolution of each 10\,kHz spectral channel within the 99--129\,MHz frequency band ranged from 23 to 30\,km\,s$^{-1}$ and 17 to 20\,km\,s$^{-1}$ at 140--170\,MHz.  Doppler correction terms are currently not incorporated into the MWA imaging pipeline; however, the velocity uncertainty is 3.4\,km\,s$^{-1}$, which is significantly less than the channel width of the MWA.  

\begin{figure}
	 \includegraphics[width=0.49\textwidth]{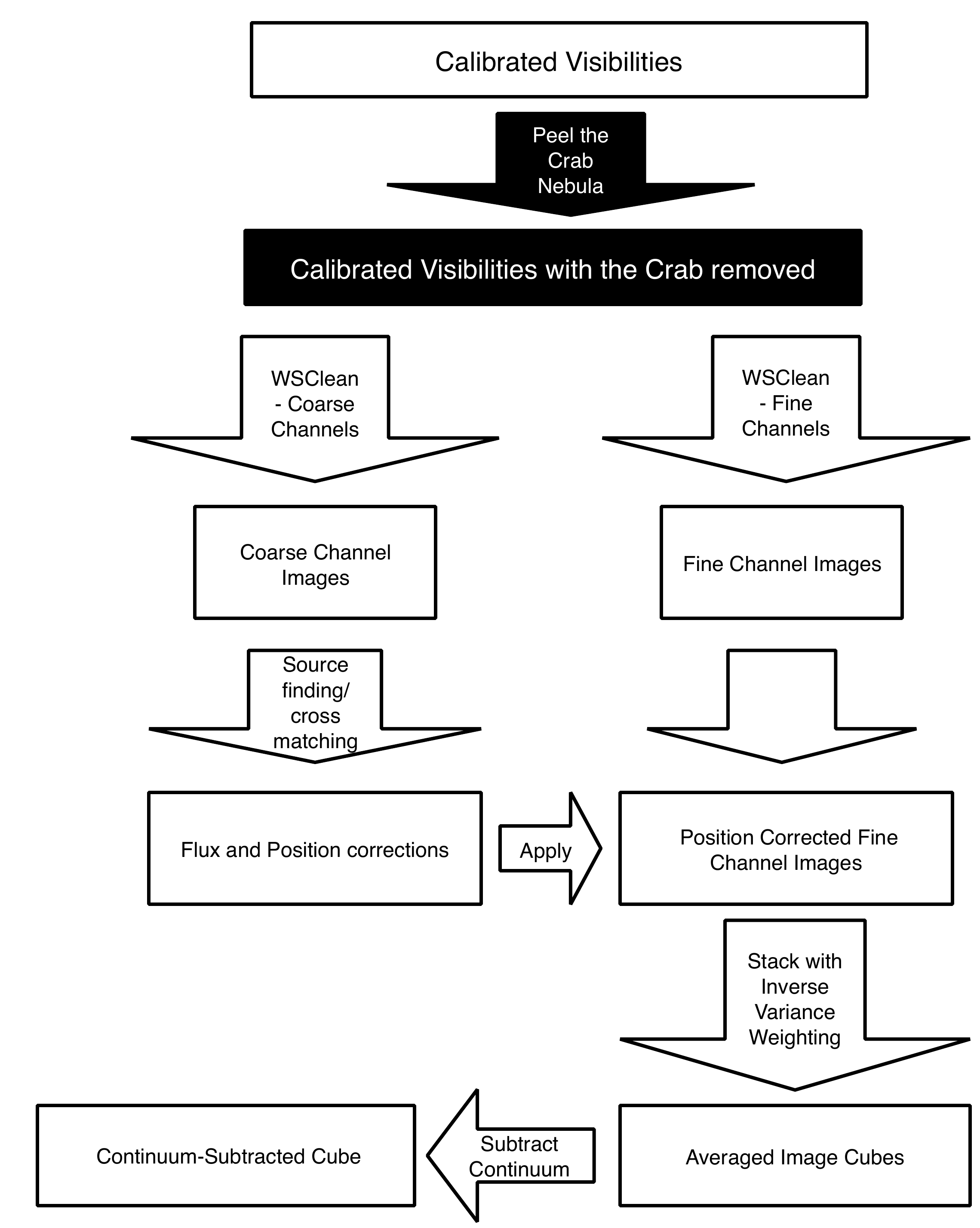}
	\caption{Summary of the pipeline designed to create calibrated time-averaged spectral image cubes with the MWA. In black are the additions to the pipeline that differ from the Galactic Center survey reported by \cite{Tremblay17}.}
	\label{pip}
\end{figure}

Data processing is completed using the pipeline published and explained by \cite{Tremblay17} (Figure \ref{pip}). However, we will highlight here some different data reduction considerations made for the Orion field.  

The MWA tiles are electronically steered so two-minute snapshot imaging was used to keep the Orion KL Nebula within the most sensitive regions of the primary beam.  Due to the large FOV of the MWA, other bright sources in the grating sidelobes of the primary beam can make a significant contribution to the image noise.  The primary difference in data processing from the Galactic Centre, shown in black in Figure \ref{pip}, is that the Crab Nebula (R.A.=05$^{h}$34$^{m}$34.94$^{s}$  Dec.=$+$22$^{\circ}$00$^{\prime}$37.6$^{\prime\prime}$, J2000; 1256\,Jy at 160\,MHz) was in the sidelobes of the primary beam. Therefore, we adopted the procedure developed during the GaLactic and Extragalactic All-sky Murchison Widefield Array (GLEAM) survey \citep{GLEAM}, to ``peel'' the Crab nebula from the visibilities.  

The compact core of the MWA increases our sensitivity to diffuse emission within our Galaxy that is intensified at low radio frequencies due to the increase brightness temperature produced by synchrotron radiation.  So even though natural weighting can be used to reduce the background noise in an image, the extended emission posed a challenge for continuum subtraction.  Therefore, natural weighting was not used to produce sky images.  Uniform weighting would provide a smaller synthesized beam, thus reducing the beam dilution of signals from compact sources.  However, by using uniform weighting we may be less sensitive to recombination lines known to trace the diffuse cold interstellar medium.  Due to the large volume of data contained within this survey, the data were not processed multiple times using different weighting schemes.  Instead the images of each coarse (1.28\,MHz) and fine (10\,kHz) channel were created using {\sc wsclean} \citep{Offringa14} with Briggs grid weighting with a robust value of ``--1'' \citep{Briggs}, a compromise between image resolution and sensitivity. The images from each coarse channel were used to check the flux density scale of observed sources in the integrated continuum image, compared to those published in the GLEAM catalog, and was found to have a mean difference of 0.1\%.

The refractive effect of the ionosphere on incident electromagnetic waves is proportional to the wavelength squared.  This causes an apparent shift in radio source positions which increases in magnitude at the lower end of our band.Therefore, a correction is built into the data processing pipeline to correct for these distortions. Single positional R.A. and Dec. offsets were applied to each fine channel of each observation based on a comparison of the source positions within the GLEAM catalog at 200\,MHz (see Figure \ref{pip}). After correction, the observations of 2015 November 22nd had mean residual positional offsets of $-$~1$\pm$17\,arc seconds in right ascension and 5$\pm$17\,arc seconds in declination.  Although the distribution is large, without the correction, the smearing of the point spread function would be larger than 10\%, impacting the source structure.  The observations on the 2015 of November 21st had less than one arc second source offsets when compared to GLEAM, so no corrections were performed.  After time-averaging all observations, the ratio of integrated flux density to peak flux density is calculated from extragalactic continuum sources within the image to be 1.01$\pm$0.07, suggesting the ionosphere is adequately corrected.  

The individual fine channels, as shown in Figure \ref{pip}, are corrected for the ionospheric effects that are derived in each coarse channel for each observation.  Due to aliasing of the polyphase filterbank, only 2400 of the 3072 fine (22$\%$) channels are imaged within the analysis pipeline. The channel images are built into a data cube and each observation was phase centered to the Orion KL Nebula prior to stacking.  This created a final integrated image of 180\,minutes for each frequency band. 

\subsection{RFI Flagging}

The observation at 140--170\,MHz, on 2015 November 21st, was significantly impacted by radio frequency interference (RFI).  Standard conditions at the observatory are radio quiet between 100--200\,MHz \citep{OffriingaRFI,Sokolowski_17} with the exception of ORBCOM at 135\,MHz, but transient RFI on November 21st affected 98\% of the data.  Therefore, only the first 2\,MHz in this frequency band is surveyed for spectral transitions.  The observations from 99--129\,MHz on 22nd November did not suffer from the same RFI problems.

RFI flagging is completed in a two-stage approach to ensure any signal that is likely due to terrestrial communication is removed.  The first stage flagged the visibility data using {\sc aoflagger} \citep{Offringa16} to remove RFI by a statistical method in each 2-minute snap-shot observation. As we would expect any astronomical sources to produce a weak signal in a 2-minute observation over 10kHz, the assumption is that any bright signals are likely associated with terrestrial communications.  {\sc aoflagger} removed <1\% of the fine channels from the individual observations at 99--129\,MHz and approximately 8\% of the channels in the observations at 140--170\,MHz.  

The RMS in the centre of most continuum-subtracted images is 0.4\,Jy\,beam$^{-1}$. To manually flag channels with spurious RFI, the image RMS is calculated for each continuum subtracted spectral fine channel within a 300$\times$300 pixel box in the middle of the fine channel continuum-subtracted image and any channel containing an RMS $>$ 0.6\,Jy\,beam$^{-1}$ is flagged\footnote{We note that by removing channels based on image RMS we may remove channels with increased spectral signals.  However, since these observations are in the same band as commercial radio, we have taken this conservative approach.}. After the two-stage RFI removal in the final cubes, the 2400 imaged channels in each frequency band are reduced to 1240 surveyed in the 99--130\,MHz data and to 200 surveyed channels in the observations at 140--172\,MHz ($\approx$55\% of the imaged bandpass).

\section{Survey Strategy}
As this survey covers a wide FOV and a large frequency range, an automated pipeline was developed to detect spectral signals in emission or absorption in the data, filter the data based on quality, and then determine if the detections are associated with known molecular transitions.  We note that the search strategy explained here is different from the pilot survey and represented the next logical progression on completing large molecular line surveys with the MWA.

The primary beam is the sensitivity pattern of the interferometer on the sky.  The 625 square degrees (Figure \ref{cont}) imaged region around Orion is reduced using {\sc mimas} \citep{Hancock12} to ensure only the most sensitive region of the primary beam (the central $\approx$65\% or down to $\approx$400 square degrees) is surveyed.  

Each of the 1440 continuum-subtracted fine-channel (10\,kHz) images are independently searched using the source-finding software {\sc Aegean} \citep{Hancock12,Hancock-2018} using the function ``slice" to set which channel in the cube is searched and a ``seed clip" value of 5, which searches the image for pixels with a peak flux density value greater than 5\,\,$\sigma$ in comparison to an input RMS image.  For the search in absorption, the option ``negative" was used to search for sources with negative flux densities.  The input RMS image is a map of the spectral RMS value at each pixel position within the continuum-subtracted data cube created for the coarse channel (100 fine channels) searched. 

Within the 95000 (3-arc\,minute FWHM) independent synthesized beams over the 400 square degree FOV and 1440 channels searched using {\sc aegean} we found 174 potential detections in emission and absorption. To publish only the most reliable detections, the catalogues of potential detections were further evaluated (or filtered) on two additional criteria; the spectral RMS and the difference between the image and spectral RMS, at the location of the spectral line detection.

The mean spectral RMS across the frequency band is 0.3\,Jy\,beam$^{-1}$ and a filter of a spectral RMS of $<$0.5\,Jy\,beam$^{-1}$ threshold is applied to the catalog of potential detections to remove the chance of false detection due to poor continuum subtraction or image artefacts.  To assess the impact of filtering the cataloged results based on the spectral RMS, we search the RMS map used in the {\sc aegean} search to find the percentage of the data impacted by this threshold.  This reduced the effective search volume by 85--90$\%$, depending on the coarse channel.

In order to simplify the statistics to regions that demonstrate Gaussian behavior, the catalogue of potential detections was further filtered to signals where the spectral signal-to-noise ratio (SNR) and image SNR both show a signal that is at least 5 and the ratio of the two RMS values (spectral and image) are within 20\% of each other. On average, this reduced the effective search volume by 74\%. 

By using two-sided Gaussian statistics at a threshold of 5\,$\sigma$, and filtering the catalogue as described above, we would expect no more than one signal to be a potential false positive.  However, we found 11 positions on the sky, with 13 signals, that passed these criteria. 


We note here that {\sc aegean} uses a Gaussian fit to the pixel data and a correction of the background to calculate the flux density for these potential detections.  However, to make the spectrum reported here, the peak flux density represents the flux density at the pixel position of potential detection as reported by {\sc aegean}.  Therefore, it is possible that the flux densities shown in each spectrum are underestimated, and their significance is greater than as listed. Our false positive rate may also be a slight overestimate as a result.

All 13 signals meeting our selection criteria were compared to known molecular and atomic transitions within the databases: Cologne Database for Molecular Spectroscopy (CDMS; \citealt{Muller}); Spectral Line Atlas of Interstellar Molecules (SLAIM; Splatalogue\footnote{www.splatalogue.net}); Jet Propulsion Laboratory (JPL; \citealt{Pickett}); and Top Model \citep{Carvajal}, for known transitions within three spectral channels ($\approx$75\,km\,s$^{-1}$ at 114\,MHz) of the detected signal.  Additionally, the source positions were searched in {\sc simbad} \citep{Wenger-SIMBAD} for plausible source emission environments, like evolved stars or nebula, within the synthesized beam.  

There are 273 known recombination and molecular transitions with upper level energy divided by the Boltzmann constant less than 300\,K (See section \S4) within the 99--129\,MHz band and 25 known transitions from 140--142\,MHz, with 58\% of the known transition contained in flagged channels.  Therefore, we would expect no more than one signal greater than 5\,$\sigma$ within the 33\,MHz observed or a 0.06\% chance any 5\,$\sigma$ signal, within $\pm$3 channels of a known transition, is random noise. When multiple transitions of a particular molecule are co-spatially located and at the same relative velocity, the chance of the signals being noise is decreased to less than 0.01\%.

If the detected signals are also associated with a known optical source that could feasibly create molecular emission, the chance that the signal is from noise is further decreased. However, we make no assumption here regarding what the line-generating source density of the region is, as no large regional molecular surveys have been performed. This is further complicated as there is a non-negligible chance that a molecular transition could be associated with a source not visible in optical or infrared surveys.   

Of the 13 signals passing our selection criteria, we found 8 that are associated with known molecular transitions, 3 are unidentified and 5 are associated with known recombination lines. In this paper we report on the molecular and unknown signals.  For information regarding the recombination lines see \cite{Tremblay-RRL}.

\section{Results \& Discussion}
A blind search, in emission and absorption, was completed as discussed in Section 3.  This corresponds to a mean flux density limit of $\approx$1.2\,Jy\,beam$^{-1}$ within the center of the primary beam. The search yielded detections of 13 signals in 11 locations. In this paper we discuss the eight signals associated with known molecular transitions and three unidentified signals.

Within the frequency bands of 99--129 and 140--141\,MHz there are 203 known molecular transitions from 18 molecules with upper state energies divided by the Boltzmann constant of 300\,K.  Although, the excitation mechanisms are not well understood at these frequencies, we consider this a reasonable limit and reduces our chances of erroneous associations.  Of the 3072 fine channels within each band, only 1440 fine channels were used across the two bands due to instrumental effects and RFI.  This reduced the number of transitions detectable in this survey by 58\%.  

Here we report tentative detections of Nitric Oxide (NO), a deuterated form of Formic Acid ($t$--DCOOH), Molecular Oxygen ($^{17}$OO), and other unidentified molecular transitions. As this is a new frequency range for this style of work, additional modeling and laboratory experiments are required to broaden the knowledge of detectable isotopolouges.

The column densities for the signals in absorption and emission are calculated using the equations in the Appendix of \cite{Tremblay17}. For each transition we use the partition function and upper level degeneracy, for low temperatures ($<$10\,K), as quoted in the CDMS catalog.  We do note that since these tentative detections have brightness temperatures of over 4000\,K, so the column densities may not be physical quantities as maser or other boosted emission mechanisms may be involved.  However, the values are provided to compare these tentative detections with detections of similar molecules in other frequency ranges.

\subsection{Nitric Oxide Tentative Detections}

Both nitrogen and oxygen are among the most common elements in the Galaxy.  Despite their relative abundance, the detection of nitric oxide (NO) is rare.  This gas-phase molecule is detected toward star-forming regions \citep{Liszt78,Gerin} and the circumstellar envelope (CSE) of evolved stars \citep{Quintana-Lacaci,Velilla-Prieto-2015,Tremblay17} with an expected formation chemistry from nitrogen atoms colliding with hydroxyl (OH) molecules \citep{Hily-Blant-2010}.  

\begin{figure}
	\centering
	 \includegraphics[width=0.52\textwidth]{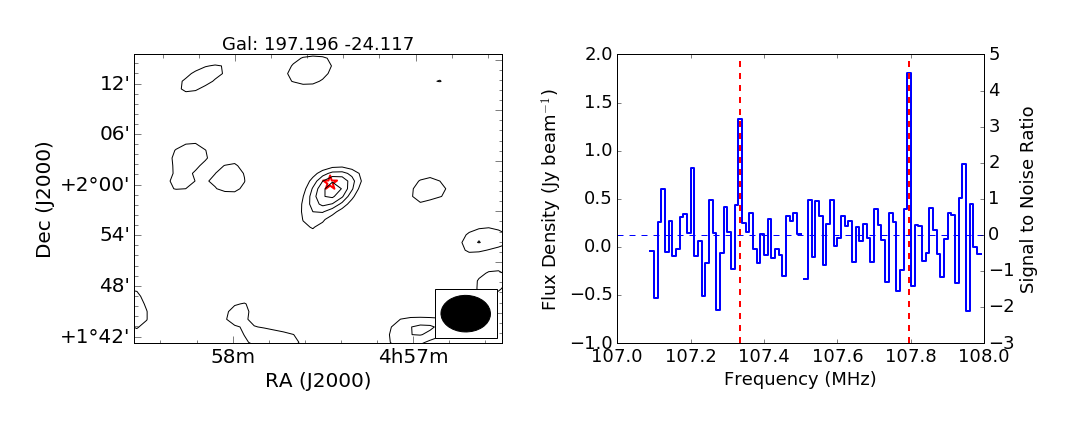}
     \centering
      \includegraphics[width=0.52\textwidth]{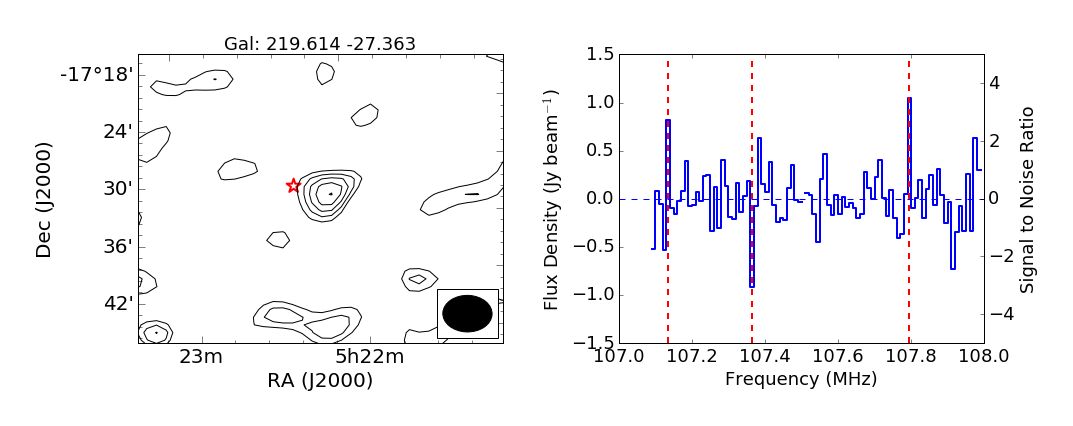}
      \centering
      \includegraphics[width=0.52\textwidth]{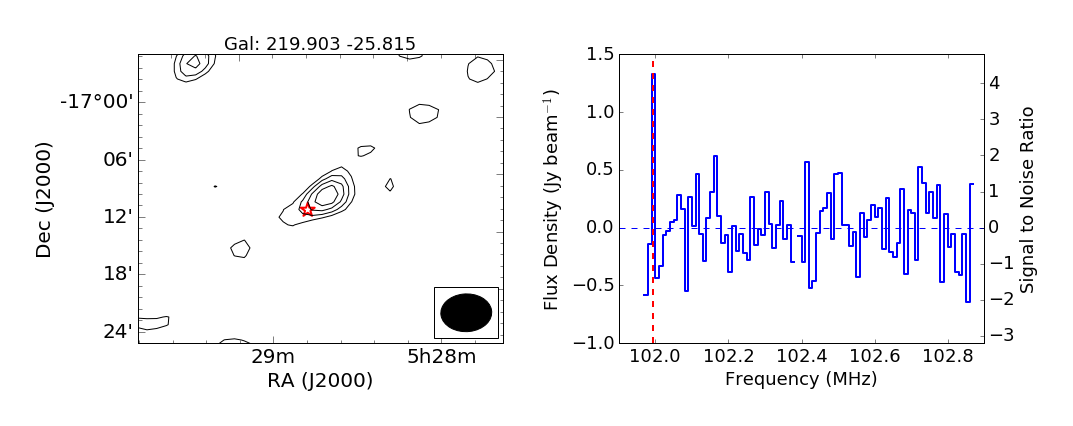}
	\caption{Tentative detections of Nitric Oxide in three stellar objects.  The spectra on the right represents the full coarse channel data cube in which the tentative detections were made.  The left panel shows the contours, set at 2, 3, 4, 5, 6\,\,$\sigma$ of the continuum-subtracted image RMS, of the detection of the brightest signal.  The ellipse in the lower right-hand corner of the contour plot shows the size and shape of the synthesized beam.  The red star is the optical position for a plausible source associated with the signal.  The size of the red star is the 1\,\,$\sigma$ astrometric error.}
	\label{NO}
\end{figure}

The spectra in Figure \ref{NO} and the data in Table \ref{NOdata} show multiple detections of NO and its isotopologues in three stellar environments which are plausible sources of the molecule.  We have tentatively detected N$^{17}$O at 107.82\,MHz and 107.17\,MHz and $^{15}$N$^{17}$O at 102.04\,MHz.  We also potentially detected a weak (3\,$\sigma$) peak of NO at 107.36\,MHz in both emission and absorption.  The three signals at G219.61~$-$27.36 are all around 3--4\,\,$\sigma$ but they are still of interest as they are all known transitions of NO at the same peak pixel position, reducing the chance that these signals are purely the result of noise.

The column density for the NO absorption peak at 107.36\,MHz is similar to the column densities of the detection of the same transition in evolved stars in the Galactic Center survey by \cite{Tremblay17} and are within the same order of magnitude of the NO column density in an evolved star detected by \cite{Quintana-Lacaci}.   

In the three stellar objects listed in Table \ref{NOdata}, no other molecular studies or detections within these objects have been published.  

\begin{table*}
\small
\caption{Information on the nitric oxide transitions tentatively and possibly detected in this survey.  The source name and type, as displayed in {\sc simbad}, is listed in the first column.  The first section of rows provides information regarding the transitions including the rest frequency, the energy in the upper level state,  the Einstein coefficient (A$_{ij}$), and the quantum numbers.  The following rows are information regarding the the peak pixel position in J2000 and Galactic coordinates and the column densities, assuming thermal emission and absorption (see \S4 for more details).}

\label{NOdata}
\begin{tabular}{llcccc}
\hline
Source	&			&	$^{15}$N$^{17}$O	&	N$^{17}$O	&	N$^{17}$O	&	NO	\\
\hline
\hline
Line Parameters	&			Rest Frequency (MHz)	&	102.04	&	107.17	&	107.82	&	107.36	\\
	&			E$_{\mathrm {U}}$/k$_{\mathrm {B}}$ (K)	&	191.15	&	191.68	&	187.32	&	7.24	\\
	&			Log$_{10}$(A$_{ij}$)	&	--16.62	&	--16.48	&	--17.30	&	--17.3	\\
	&			Quantum Numbers	&	J=$\frac{5}{2}$--$\frac{5}{2}$ 	&	J=$\frac{3}{2}$--$\frac{3}{2}$, 	&	J=$\frac{15}{2}$--$\frac{15}{2}$, 	&	J=$\frac{3}{2}$--$\frac{3}{2}$, 	\\
	&				&	p=-1--1 	&	$\Omega$=$\frac{3}{2}$ 	&	$\Omega$=$\frac{3}{2}$ 	&	$\Omega$=$\frac{1}{2}$ 	\\
	&				&	F$_{1}$=3--4, F$_{2}$=3--4	&	F$_{1}$=$\frac{5}{2}^{+}$--$\frac{7}{2}^{-}$, F$_{2}$=$\frac{3}{2}^{+}$--$\frac{5}{2}^{-}$	&	F$_{1}$=$\frac{13}{2}^{+}$--$\frac{15}{2}^{-}$, F$_{2}$=$\frac{15}{2}^{+}$--$\frac{17}{2}^{-}$	&	F=$\frac{5}{2}^{+}$--$\frac{3}{2}^{-}$	\\
\hline
CRTS J0457275 +015840			&	Position (J2000)	&		&		&	4h57m32s +01$^{o}$56$^{\prime}$53$^{\prime}$$^{\prime}$&		\\

W UMA Eclipsing Binary	&			Position (Galactic)	&		&		&	197.196 --24.117	&		\\
	&			Velocity (km\,s$^{-1}$)	&		&		&	--56$\pm$28	&	--28$\pm$28	\\
	&			N$_{u}$ (cm$^{-2}$)	&		&		&	1.2$\times$10$^{21}$	&	9.6$\times$10$^{20}$	\\
	&			N$_{tot}$ (cm$^{-2}$)	&		&		&	1.1$\times$10$^{23}$	&	1.0$\times$10$^{14}$	\\
    \hline
UCAC2 25008117			&	Position (J2000)	&		&		&	5h22m12s -17$^{o}$33$^{\prime}$02$^{\prime}$$^{\prime}$	&		\\
Evolved Star	&			Position (Galactic)	&		&		&	219.614 --27.363	&		\\
	&			Velocity (km\,s$^{-1}$)	&		&	--56$\pm$28	&	--56$\pm$28	&		0$\pm$28\\
	&			N$_{u}$ (cm$^{-2}$)	&		&	6.9$\times$10$^{20}$	&	9.5$\times$10$^{20}$	&		\\
	&			N$_{tot}$ (cm$^{-2}$)	&		&	3.3$\times$10$^{23}$	&	8.5$\times$10$^{22}$	&	9.0$\times$10$^{20}$	\\
\hline
BD-17 1144			&	Position (J2000)	&	5h28m37s -17$^{o}$12$^{\prime}$21$^{\prime}$$^{\prime}$	&		&				\\
Evolved Star	&			Position (Galactic)	&	219.903 -25.815	&		&		&		\\
	&			Velocity (km\,s$^{-1}$)	&	--94$\pm$32	&		&		&		\\
	&			N$_{u}$ (cm$^{-2}$)	&	1.6$\times$10$^{21}$	&		&		&		\\
	&			N$_{tot}$ (cm$^{-2}$)	&	4.1$\times$10$^{23}$	&		&		&		\\
						
\hline
\end{tabular}
\small{All transition values are from the CDMS -- Cologne Database for Molecular Spectroscopy Database \citep{Muller}}
\end{table*}

\vskip 0.8in

\subsection{Formic Acid Tentative Detection}
The smallest of the organic acids, formic acid, is an abundant gas-phase molecule \citep{Bisschop-2007} observed in regions of star formation and quiescent clouds \citep{Linnartz, Bennett-2011}.  While the formation routes are debated, laboratory experiments have shown that formic acid can form through hydrogenation of the H--O C--O complex at low temperatures in oxygen rich environments \citep{2011MNRAS.413.2281I}.  

\begin{figure}
	\centering
	 \includegraphics[width=0.52\textwidth]{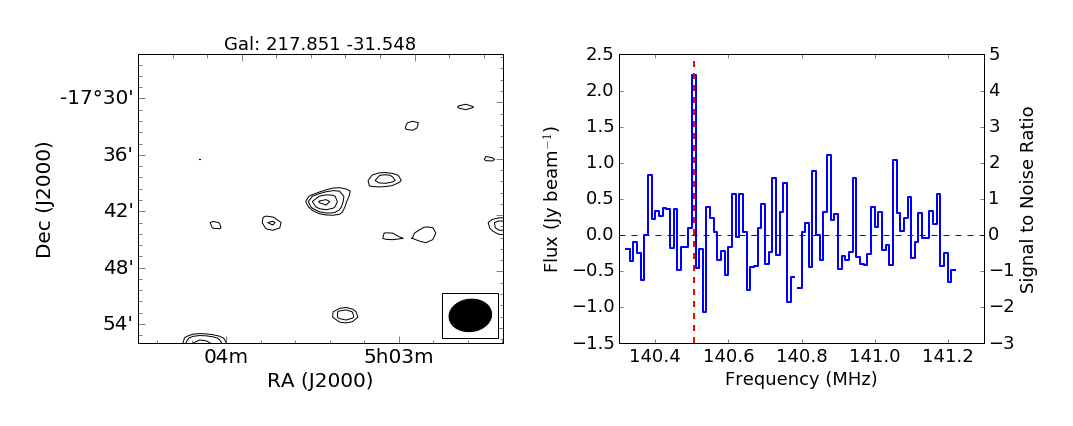}
	\caption{Tentative detections of Formic Acid with the right panel showing the spectra of the full coarse channel data cube containing the tentative detection.  The left panel is the contours, set at 2, 3, 4, 5, and 6\,\,$\sigma$ of the image RMS, of the detected signal in the continuum subtracted image. The ellipse in the lower right-hand corner of the contour plot shows the size and shape of the synthesized beam.}
	\label{Formic}
\end{figure}

We have tentatively detected a singly deuterated form of formic acid ($t$--DCOOH) where one of the hydrogens is replaced with deuterium (also known as heavy hydrogen or $^{2}$H), a stable isotope of hydrogen formed during Big Bang Nucleosynthesis. The spectra of the tentative detection at 140.5\,MHz (Figure \ref{Formic}) is not associated with any known source contained in {\sc simbad} but is located within the Orion molecular cloud. Deuterated species are expected to be enhanced in cold regions of early star formation \citep{Neill-2013}, so this may not be surprising.

Theory suggests the ratio of D/H is $\approx$10$^{-5}$ \citep{Lacour-2004} but \cite{Neill-2013} found regions of enhanced D/H ratio in the Orion molecular ridge to be $\approx$5$\times$10$^{-3}$.  These enhanced regions D/H are correlated with the first cold prestellar phase of star formation \citep{Skouteris-2017} because of the CO depletion onto the surface of dust grains which makes deuterated molecules important for studying the dynamics of molecular clouds \citep{Das-2015}. 

The column density, in Table \ref{Formdata} of 3.6$\times$10$^{20}$ is higher than other deuterated species calculated by \cite{Neill-2013} with the difference possibly due to the boosted emission likely detected at low frequencies \citep{SpitzerISM}.  One other known transition of this molecule within our frequency band is at 124.80\,MHz and is a channel flagged in these observations. 

\begin{table}
\small
\caption{Information on the deuterated formic acid transition tentatively and possibly detected within this survey.  The first row provides information regarding the transition including the rest frequency, the energy in the upper level state, the Einstein coefficient (A$_{ij}$), and the quantum numbers.  The proceeding rows are information regarding the the peak pixel position in J2000 and Galactic coordinates, the velocity of the transition, and the column densities, assuming thermal emission and absorption (See \S4 for more details).}

\label{Formdata}
\begin{tabular}{llcccc}
\hline
Source	&	Unknown\\
\hline
\hline
Molecule	&	$t$-DCOOH\\
Rest Frequency (MHz)	&	140.50\\
E$_{\mathrm {U}}$/k$_{\mathrm {B}} (K)$	&	200.16\\
Log$_{10}$(A$_{ij}$)	&	--14.24\\
Quantum Numbers	&	16(5,11)--16(5,12)\\
Position (J2000)	&	5h03m30s --17$^{o}$40$^{\prime}$28$^{\prime}$$^{\prime}$\\
Position (Galactic)	&	219.903 --25.815\\
Velocity (km\,s$^{-1}$)	&	0$\pm$17\\
N$_{u}$ (cm$^{-2}$)	&	1.1$\times$10$^{19}$\\
N$_{tot}$ (cm$^{-2}$)	&	3.6$\times$10$^{20}$\\					
\hline
\end{tabular}
\\
\small{All transition values are from the JPL -- Jet Propulsion Laboratory Spectral Line Catalog \citep{Pickett}}
\end{table}

\subsection{Molecular Oxygen Tentative Detection}
Oxygen is the most abundant heavy element in our Galaxy and is expected to be primarily tied up in CO, O$_{2}$ and H$_{2}$O, with O$_{2}$ following closely with the abundance of CO \citep{Goldsmith-2011}. Due to the water in the troposphere around Earth, the detection of O$_{2}$ and H$_{2}$O within the Milky Way at high frequencies is difficult with millimeter-wave ground based telescopes.  However, at low radio frequencies the Earth's atmosphere is optically thin and so detections of oxygen transitions become viable.  \cite{Goldsmith-2011} provides a review of detections of molecular oxygen in the interstellar medium and laboratory experimentation as well as reporting detections in the Orion molecular cloud with the $Herschel$ $Space$ $Observatory$.  Their detections of molecular oxygen at 487, 774 and 1121\,GHz had velocities of 11 to 12\,km\,s$^{-1}$ with line widths of 3\,km\,s$^{-1}$ and column densities averaging around 5$\times$10$^{16}$\,cm$^{-2}$.  

The formation pathway of O$_{2}$ and its isotopologues is thought to be initiated by the reaction of atomic oxygen and H$_{3}$$^{+}$, which is produced through the ionosation of H$_{2}$ from cosmic rays \citep{Goldsmith-2011}.  The recent calculation of reaction rates suggest that large O$_{2}$ production in the gas phase is possible \citep{2009JChPh.131v1104L} and \cite{Goldsmith-2011} calculated that the three transitions they detected came from a source of 5\,arc\,seconds in angular size, 10--11\,M$_{\odot}$, and not associated with any sources of previous molecular detections.

We have tentatively detected molecular oxygen ($^{17}$O$^{16}$O) at a level of 5\,\,$\sigma$ in comparison to the spectral RMS.  The asymmetric nature of the ($^{17}$O$^{16}$O) molecule gives it a dipole moment, allowing it to be detected at radio frequencies.  \cite{Pagani-1993} made the first tentative radio detection of molecular oxygen ($^{16}$O$^{18}$O) in the stacked spectra of interstellar cloud NGC7538 and L134N with a calculated integrated intensity of 13 and 5\,mK\,km\,s$^{-1}$, respectively.  However, these detections were never replicated \citep{Goldsmith-2011}. 

Within our observations there are 34 transitions of $^{17}$OO but 18 of them are flagged out of our observations.  The detected signal at 107.60\,MHz is one of the lowest energy transitions.  At 111.03\,MHz the transition with the quantum numbers N=3--3, J=2--2, F=$\frac{7}{2}$--$\frac{5}{2}$ has an E$_{u}$/k$_{\mathrm{b}}$ of 23.7\,K, but was not detected within this survey.  

\begin{figure}
     \centering
      \includegraphics[width=0.52\textwidth]{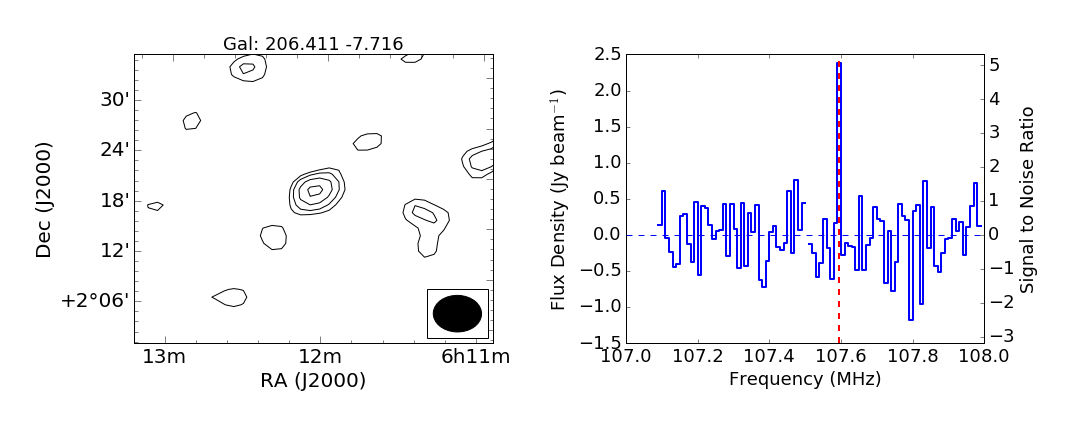}
     	\caption{Tentative detections of Molecular Oxygen. The right panel shows the spectra of the full coarse channel data cube containing the tentative detection.  The left panel is the contours, set at 2, 3, 4, 5, and 6\,\,$\sigma$ of the image RMS, of the detected signal in the continuum subtracted image.  The ellipse in the lower right-hand corner of the contour plot shows the size and shape of the synthesized beam.}
	\label{OO}
\end{figure}

The detection, shown in Figure \ref{OO}, is not at a location of any known source listed in {\sc simbad}.  The column density in Table \ref{Oxdata} of 1.3$\times$10$^{25}$\,cm$^{3}$ is higher than the $Herschel$ detections, but that could be due to the non-thermal emission or our incorrect assumption on the rotational temperature (assumed 9\,K for this work).

\begin{table}
\small
\caption{Information on the molecular oxygen transition tentatively and possibly detected within this survey.  The first row provides information regarding the transition including the rest frequency, the energy in the upper level state, the Einstein coefficient (A$_{ij}$), and the quantum numbers.  The proceeding rows are information regarding the the peak pixel position in J2000 and Galactic coordinates, the velocity of the transition, and the column densities, assuming thermal emission and absorption (See \S4 for more details).}

\label{Oxdata}
\begin{tabular}{llcccc}
\hline
Source	&	Unknown\\
\hline
\hline
Molecule	&	$^{17}$O$^{16}$O\\
Rest Frequency (MHz)	&	107.60\\
E$_{\mathrm {U}}$/k$_{\mathrm {B}}$	(K)&	39.81\\
Log$_{10}$(A$_{ij}$)	&	--18.01\\
Quantum Numbers	&	N=4--4, J=3--3, F=$\frac{7}{2}$--$\frac{5}{2}$\\
Position (J2000)	&	6h12m04s +02$^{o}$13$^{\prime}$54$^{\prime}$$^{\prime}$\\
Position (Galactic)	&	206.411 $+$7.716\\
Velocity (km\,s$^{-1}$)	&	0$\pm$17\\
N$_{u}$ (cm$^{-2}$)	&	1.2$\times$10$^{24}$\\
N$_{tot}$ (cm$^{-2}$)	& 1.3$\times$10$^{25}$	\\					
\hline
\end{tabular}
\\
\small{All transition values are from the JPL -- Jet Propulsion Laboratory Spectral Line Catalog \citep{Pickett}}
\end{table}

\subsection{Unidentified Tentative Detections}
\begin{figure}
	\centering
	 \includegraphics[width=0.52\textwidth]{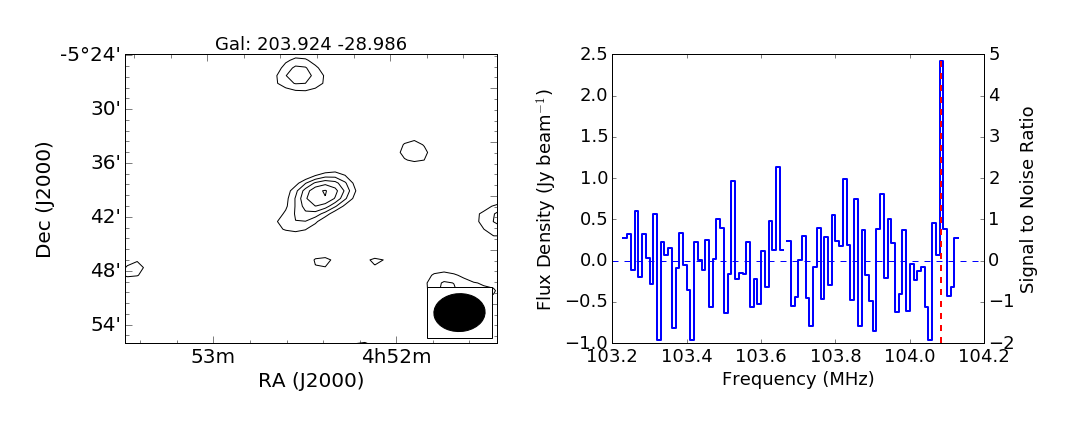}
      \centering
      \includegraphics[width=0.52\textwidth]{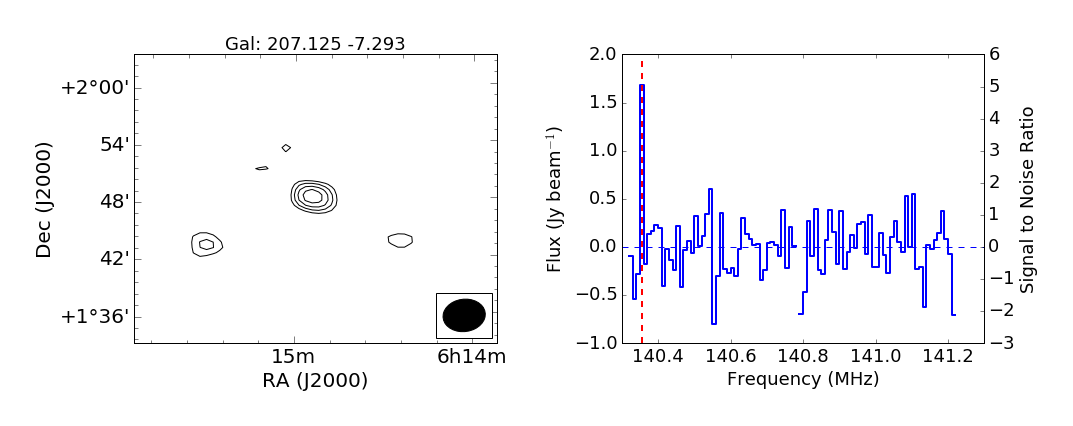}
      \centering
      \includegraphics[width=0.52\textwidth]{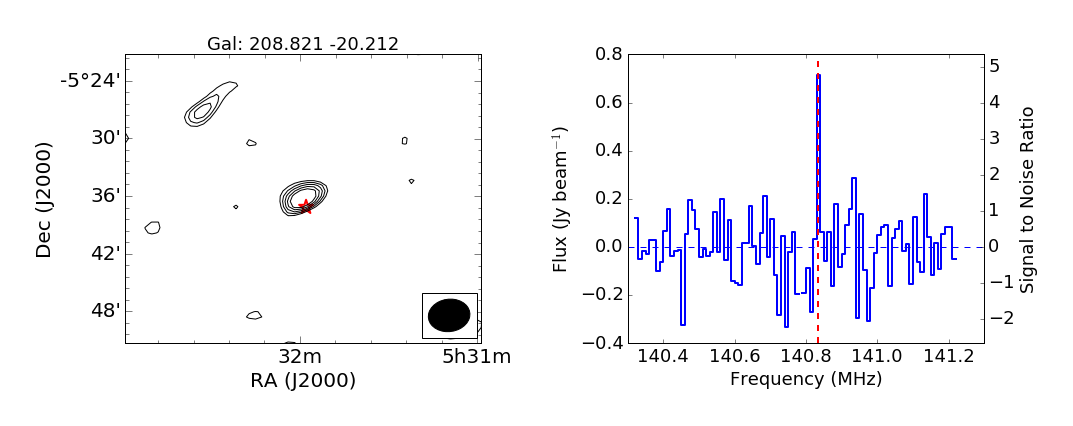}
	\caption{Tentative detections of Unknown Molecules.  The right panel shows the spectra of the full data cube containing the tentative detection.  The left panel is the contours, set at 2, 3, 4, 5, and 6\,\,$\sigma$ of the image RMS, of the detected signal in the continuum subtracted image. The ellipse in the lower right-hand corner of the contour plot shows the size and shape of the synthesized beam. The signal at G208.821 -20.212 is located within the Nebula with the closest optical source at the peak pixel position, marked with a red star, is Parenago 811 an evolved star within the Nebula.}
	\label{Unknown}
\end{figure}

We have tentatively detected three signals at around 5\,$\sigma$ in comparison to the spectral RMS that are not associated with any known molecular or recombination line transitions at 104.09, 140.36, and 140.84\,MHz (Figure \ref{Unknown}).  Neither of the first two transitions are co-located with a known object listed in {\sc simbad} within the area of our synthesized beam.  However, the signal at 140.84\,MHz is located within the Orion KL region and is most closely associated with the evolved star Parengo 811.

\subsection{Orion Chemical Environment}
\cite{Baudry-2016} provides a summary of the complex chemical environment of the Orion molecular cloud with a particular focus on high resolution information provided by the Atacama Large Millimeter/submillimeter Array (ALMA) and IRAM 30\,m telescopes.  They state that although most of the interstellar molecules detected can be found in Orion, there is a dominance of nitrogen and oxygen complex molecules.  The $Herschel$ $Space$ $Observatory$ with the HIFI detector has also provided a wealth of detailed information regarding the chemical complexity of Orion over recent years (e.g. \citealt{Goldsmith-2011,Neill-2013,Crockett}), in particular with detection of molecules like molecular oxygen which can be difficult with ground based telescopes.

Most of the molecules detected around Orion KL Nebula or the molecular cloud have velocities $<$15\,km\,s$^{-1}$ and line widths around 3\,km\,s$^{-1}$.  The resolution of the MWA of 17--30\,km\,s$^{-1}$ suggests that all the detections would be within one channel and close to the rest frequency, which is what is observed for molecular oxygen and formic acid.  However, the nitric oxide and nitric oxide isotopologue detections have a higher velocity and are possibly associated with evolved stars, which are known to have velocities up to 150\,km\,s$^{-1}$ \citep{Olofsson}.  \cite{Velilla-Prieto-2015} observed several transitions of NO at 150 to 350\,GHz in the CSE of OH231.8$+$4.2, with velocities of 35--40\,km\,s$^{-1}$ and line widths up to 50\,km\,$^{-1}$, which is consistent with the expectation of detecting broad velocity components with the MWA due to our spectral resolution.

\subsection{Connections to Laboratory Science}

\cite{Fedoseev-2012} confirmed that NO plays an important role in the formation of hydroxylamine (H$_{3}$NO), which is an important molecule in the pathway of the formation of amino acids.  Laboratory experiments have created enamines (precursors to amines) with deuterated Formic Acids in a neutral medium \citep{Himmele-1979}, suggesting that the tentative detections of deuterated formic acid and nitric oxide may point to the production of amines within the Orion molecular cloud.  NO is also of particular interest as it is thought to be critically important in primitive life on Earth \citep{feelisch_martin_1995, Santana-2017}.

The detection of signals not associated with known molecules or known position of sources is not unexpected.  As the study of molecules within the frequency range of 100--170\,MHz is new, not a lot of emphasis has been put into publishing transitions at these frequencies.  Additionally, the wide FOV of the MWA is unique, giving us the opportunity to study much larger regions than previously available in the radio part of the electromagnetic spectrum.  Many molecular studies are completed in regions where interesting sources are already known to exist instead of blind surveys. Additional laboratory experiments at these frequency ranges would allow us to better identify potential molecules and start to understand the abundances of the molecules we are detecting. 

\section{Conclusions}
We completed a survey from 99--170\,MHz centered on the Orion KL nebula to search for molecules. Due to unusual RFI and instrumental effects only 23\% of the channelized data were available to search for absorption and emission signals likely associated with molecular transitions.  Despite this we have tentatively detected multiple transitions of nitric oxide, deuterated formic acid, and molecular oxygen.  We also have three tentative detections of signals that have no known molecular or recombination line associations.  We call these tentative as they are all around 5\,$\sigma$ and are contained within one channel.  

With an integration time of three hours and the wide velocity width of the MWA, we are likely only sensitive to boosted emission with broad velocity components.  However, our results are consistent with our Galactic plane survey \citep{Tremblay17} and theoretical assumptions of the nature of the transitions we would expect at low-frequencies \citep{SpitzerISM,CondonERA}.  Although the individual tentative detections represent low significance, the fact that we see multiple transitions of nitric oxide within the same object is promising.      

Molecules are often used as classification tools for identifying and studying environments in greater detail than continuum observations can provide. Although the spectral resolution of the MWA does not allow us to study the kinematics of the regions, the MWA's wide FOV is well suited to finding new and interesting molecular environments that can be followed up with high-resolution telescopes. In this survey we detected deuterated formic acid; deuterated molecules are known to be enhanced during the early stages of star formation.  Therefore, future observations with the MWA, using a longer integration time to reduce the thermal noise, will be useful for identifying regions for further study to understand the early stages of star formation.

The detections of both molecular oxygen and deuterated formic acid point to the possible formation of amines within the Orion complex.  As the fundamental transitions of larger molecules like amines reside at low radio frequencies, they are ideal frequencies for searching for these missing building blocks in molecular evolution.

The MWA has recently undergone an upgrade to longer baselines and new software is being designed to improve the spectral quality and channel resolution.  These together with increase integration time will allow for improved blind surveys in the future.  Additionally, all of these are in preparation for understanding requirements for the low-frequency Square Kilometre Array (SKA) planned to be built on the same site as the MWA.

Future work with the MWA is planned to use the new long baselines and increased integration times.  As found by \cite{Tremblay17}, it will take approximately 30 hours of observation to get to the sensitivity limits to find new or rare molecules.  Therefore, future observations are in progress to observe the Vela Molecular Ridge and Gum Nebula for over 30 hours.

This work reported here further validates the feasibility of using low-frequency radio telescopes to complement the work at high-frequencies.  Both ranges have their own strengths and low-frequencies can provide insights into environments as they probe different conditions (eg.
non-thermal excitation). 

\acknowledgments
We would like to thank the referee for their insightful comments and time to improve our manuscript. The authors would like to acknowledge the contribution of an Australian Government Research Training Program Scholarship in supporting this research. This work was supported by resources provided by the Pawsey Supercomputing Centre with funding from the Australian Government and the Government of Western Australia. This scientific work makes use of the Murchison Radio-astronomy Observatory, operated by CSIRO. We acknowledge the Wajarri Yamatji people as the traditional owners of the Observatory site. Support for the operation of the MWA is provided by the Australian Government (NCRIS), under a contract to Curtin University administered by Astronomy Australia Limited.

\facility{MWA}
\software{AEGEAN \citep{Hancock12,Hancock-2018}, WSCLEAN \citep{Offringa14}, AOFLAGGER \citep{Offringa16}, COTTER \citep{Offringa15}}, TOPCAT \citep{Topcat}

\bibliographystyle{yahapj}
\bibliography{research3}


\end{document}